# Large Field, High Resolution Full Field Optical Coherence Tomography: A Pre-clinical study of human breast tissue and cancer assessment

Osnath Assayag[1], Martine Antoine[2], Brigitte Sigal-Zafrani[3], Michael Riben[4], Fabrice Harms[5], Adriano Burcheri[1], Bertrand Le Conte de Poly[5] and A. Claude Boccara[1,5]

**Authors Affiliations:** [1]Institut Langevin ESPCI ParisTech, Paris, France ; [2] Department of Pathology, Tenon Hospital APHP, Paris, France; [3] Department of Tumor Biology, Institut Curie, Paris, France; [4] Department of Pathology, University of Texas MD Anderson Cancer Center, Houston, Tx, USA; [5] LLTech SAS, Paris, France

**Corresponding Author:** Osnath Assayag, Institut Langevin, ESPCI ParisTech, 10 rue Vauquelin, 75005 Paris. Phone: +33 1 40 79 45 83. E-mail : osnath.assayag@espci.fr

**Abstract:**

We present a benchmark pilot study in which high-resolution Full-Field Optical Coherence Tomography (FF-OCT) is used to image human breast tissue and is evaluated to assess its ability to aid the pathologist's management of intra-operative diagnoses. Our aim included evaluating the safety of FF-OCT on human tissue and determining the concordance between the images obtained with routinely prepared histopathological material.

The compact device used for this study provides a 1.5 µm-lateral and a 1 µm-axial resolution, and is able to scan a 1.5cm² specimen in about 7 minutes. 75 breast specimens obtained from 22 patients have been imaged. Because the contrast in the images is generated by endogenous tissue components, no biological or chemical contrast agents or specimen preparation are required.

We characterized the major architectural features and tissue structures of benign breast tissue, including adipocytes, fibrous stroma, lobules and ducts. We subsequently characterized features resulting from their pathological modification and developed a decision tree for diagnosis. Two breast pathologists applied these criteria, resulting in a demonstrable ability to distinguish between normal or benign tissue, in situ and invasive carcinomas using FF-OCT images, with a sensitivity of 97% and 90%, respectively, and specificity of 74% and 77% respectfully.

FF-OCT shows great potential for the evaluation of human tissue and its characterization as normal/benign vs. lesional, for numerous ex-vivo clinical use-cases. Its high imaging accuracy for in-situ and invasive carcinoma paves the way for applications where a fast architectural assessment could improve the core needle biopsy workflow and tumor margin assessment.

**Introduction:**

Breast cancer remains the second leading cause of cancer death in women in the United States, with an estimated 232,620 new cases and 39,970 breast cancer deaths (39,520 women, 450 men) expected in 2011 (1). Nevertheless, in many countries, thanks to more systematic screening and better treatment regimens, breast cancer mortality has steadily decreased since 1990, with a decrease of 3.2% per year in women younger than 50 and of 2% per year for women 50 and older (1). Earlier screening allows for the detection of small invasive lesions and in situ carcinomas. For these lesions, breast-conserving surgery (lumpectomy) is standard practice, along with axillary sentinel node biopsy, in cases of invasive carcinoma.

Prior to definitive curative surgical therapy, routine diagnostic evaluation, usually with needle core biopsy and/or fine needle aspiration, has obviated the need for lesional diagnostic frozen section analysis at the time of surgery (2). In fact, for very small tumors, some have suggested that frozen sections may even be inappropriate for lesional diagnosis (3). However, frozen section analysis may still be performed to assess surgical margins, in an effort to minimize the need for a second surgical procedure. Unfortunately, frozen section analysis reportedly only has an overall sensitivity of 73% (4-5) and is associated with several drawbacks: it is time consuming (20-30min) (6), highly operator dependent, destructive to the sample as part of the tissue is lost during slide preparation, allows for only limited sampling, and introduces freezing artifacts (7-11) that significantly hamper interpretation. Alternative techniques for immediate assessment of specimen margins, such as touch prep cytology (12) or intraoperative radiography (13), are also fraught with disadvantages, including poor sensitivity, limited spatial resolution, and added time. As a result, the re-excision rate is typically 20%, and may be as high as 40% (5-6, 14-19). Consequently, improvements in the ability to perform accurate intra-operative margin assessment using a sub-surface optical microscopic technique is attractive, particularly if it is easy to perform, operator independent, non-destructive, fast, and provides for high enough spatial resolution that can mimic traditional histopathological analysis.

Successful optical microscopy has been achieved on excised skin (Mohs surgery) using confocal microscopy (20) and multiphoton microscopy (21). However, the need for large numerical aperture objectives to "virtually slice" the tissue makes confocal microscopy sensitive to tissue-induced aberrations as well as to the effects of multiple scattering. Alternatively, optical sectioning techniques using Optical Coherence Tomography (OCT) (22) with a deep red or near infrared sample illumination offers many advantages: they are safe, non destructive (no biological or chemical modification), and do not require any exogenous contrast agents that could potentially compromise the tissue integrity for subsequent routine histopathological preparation or molecular testing. In the current study, we use the Full-Field OCT (FF-OCT)(23-24) technique to provide large fields of view and high resolution images of unprocessed breast specimens.

*Comparison of FF-OCT to other current imaging techniques*

The principal imaging techniques comparable to FF-OCT are traditional OCT (22), confocal microscopy (25), and Optical Coherence Microscopy (OCM) (26).

Numerous studies have been published during the past few years, demonstrating the suitability of OCT imaging for both *in vivo* and *ex vivo* tissue diagnosis. OCT has been previously investigated in a variety of tissues such as the eye (27-29), upper aerodigestive tract (30-33), gastrointestinal tract (34), breast tissue and lymph nodes (35-41). One of the earliest studies on human breast tissue showed that benign and malignant lesions could be differentiated using "Ultrahigh Resolution and Three-dimensional Optical Coherence Tomography" (37), which correlated OCT images with histopathological findings for ductal carcinoma in situ, invasive cancer and micro-calcifications. Another study (38), based on spectral-domain OCT, focused on the assessment of breast tumor margins and demonstrated good sensitivity and specificity. However, this study, involving a significant number of patients and images, is specific to tumor margin evaluation and uses diagnostic criteria based on "large scatterers" and not on the microscopic morphology. Both these studies of breast tissue utilized techniques that had a limited spatial resolution: the former obtained a 6 μm transverse resolution and 3.5 μm axial resolution, while the latter obtained a 6μm axial resolution and 35μm transverse resolution. Unfortunately, at these resolutions, OCT images appear fuzzy compared to conventional histology. Indeed, in order to satisfy pathologist's requirements, the optical technique must mimic that seen with traditional histology : typically a field of view must be larger than a $cm^2$ and the smallest details (spatial resolution) should be closer to 1 μm.

Technically , the FF-OCT setup, described in details in the methods section below, uses immersion microscope objectives with a numerical aperture of 0.3. This is in contrast to typical OCT setups in which the numerical aperture is one order of magnitude lower (42) therefore the light scattered by small sized structures smaller than the wavelength is between 10 (for fiber type ones) and 100 (for 3-D structures times weaker and are missing in low aperture images; yielding OCT images with resolution and contrast that is comparatively much lower than FF-OCT images.

Furthermore, as it is described in the following section, FF-OCT uses "en face" acquisition of the images. Due to this "en face" acquisition , FF-OCT offers by far the best lateral resolution. Indeed FF-OCT can work without the large depth of field required for time domain, spectral domain, Fourier-domain or swept-source OCT. For these non FF-OCT methods, the available depth range must be of the same order as the depth of field of the objective, therefore requiring low numerical aperture optics, which consequently limits the lateral resolution capability to typically 5 to 40 μm. In contrast, the lateral resolution of FF-OCT is comparable to Optical Coherence Microscopy (24), but its axial resolution is better than what could be obtained using super-fast lasers, while facilitating more flexibility and a better ability to capture a large field of view.

Improved resolution has recently been accomplished by combining two techniques: OCT, with its large field of view, and Optical Coherence Microscopy (OCM) with its maximum lateral resolution of 2 μm (26). By doing so, it was possible to achieve high-resolution images of small zones with a sectioning ability (axial resolution) of 4 μm. With this improved resolution, the comparison of images with conventional histology of breast tissue is striking, providing an ability to distinguish adipose tissue, fibrous stroma, breast lobules and ducts, as well as in-situ and invasive carcinomas. For that study, the field of view of OCM was limited to $400 \times 400 \mu m^2$, utilizing a femtosecond ND:glass laser The large field images can only be obtained using the low axial (4μm) and transverse (20μm) resolution of the OCT system; at this resolution breast structures appear blurry. High resolution images can only be obtained by switching to OCM which then limits the field of view to $400 \times 400 \mu m^2$. Other limitations to the study included the inability to carry out accurate statistical analysis of sensitivity and specificity due to few specimens and an inability to determine detection accuracy of the new imaging technology due to non-blinded evaluation of the images.

Pathologists need to be able to identify a zone of interest for assessing the malignancy of a breast specimen. However, with the low resolution of the large field image it can be difficult to identify the correct zone of interest and the tiny field of view when using OCM represents a major drawback for implementation in the clinical setting.

In addition, the cost of an adequate femtosecond laser remains very high. Although laser technology continues to improve, these lasers remain expensive, often costing as much as the rest of the OCM setup. Large peak power at the focus of ultra fast laser can induce high order photobleaching, similar to what has been observed using 2-photon fluorescence (43). Other endogenous chromophores, such as melanin can limit tissue imaging by causing thermal and mechanical damage during an experiment. Finally using lasers requires supplementary safety requirements. For these reasons, OCM is not the best method to use for all tissue imaging application.

Confocal microscopy (CM) produces high resolution images but the imaging depth is markedly limited. Whereas FF-OCT can show clear signals (contrast and resolution) down to 200µm in depth for fat and 150µm in depth for fibrous tissue, the depth is limited to about 50µm with a confocal microscope.

FF-OCT allows a pathologist evaluating large fields to scan the entire biopsy specimen and to follow the morphology at different scales and different positions. The Full-Field OCT (FF-OCT) technique accommodates imaging of unprocessed tissue samples down to 200 µm below the specimen surface (depending on the type of tissue), over a large field of view that allows zooming in on specific areas down to the micron scale. More importantly, it should be emphatically noted that no exogenous contrast agent nor any chromophore is required for imaging, as specimens are excised from mastectomy pieces and can be analyzed fresh without fixative and directly imaged with no further preparation. Utilization of FF-OCT for imaging of both fresh and fixed animal tissue has already proven its suitability and performance for high-resolution, large field images (44).

The main objective of this study is to evaluate whether the images obtained with FFOCT contain sufficient morphologic details that is adequate for a pathologist to make a suitable diagnosis, comparable to that which would be rendered using traditional histological techniques. Specifically, to assess the ability to identify various tissue structures, the ability to assess if an area is "normal" or contains "pathology", and the ability to identify specific histological patterns characteristic of common pathological entities.

We aim to show in the present study that tissue characterization is now possible.

**Material and Methods:**

*Instrument:*

OCT (22) is a unique technique for imaging scattering media such as biological tissues, by interferometric selection of ballistic photons. It has proved to be an invaluable tool for biomedical imaging. In contrast with most of the available OCT approaches (e.g. time domain OCT or Fourier domain OCT), FF-OCT directly takes "en face" images using megapixel cameras and microscope objectives. In our setup, a very compact and stable interferometer is displaced to move the focal plane at different depths below the surface and to generate 3D tomographic images(45). En face capture allows FF-OCT to operate with high lateral resolution (typically ~1µm) using medium to large aperture microscope objectives (46). Nevertheless, the penetration of the beam into the volume of the sample induces a shift between the focus and the sectioning plane due to refractive index mismatch. In order to compensate for this phenomenon and the aberrations that may take place, a real time optimization of the signal is achieved (47).

The principle of FF-OCT relies on low coherence interference microscopy. The experimental arrangement of FF-OCT (Figure 1) is generally based on a configuration that is referred to as a Linnik interferometer (48). A halogen lamp is used as a spatially incoherent source to illuminate the whole field of a immersion microscope objective. Due to the low temporal coherence of the source, interference occurs only when the optical path lengths of the two-interferometer arms are identical within typically 1µm, which is equal to the effective coherence length of the source (taking into account the camera spectral response) divided by $2n$, $n$ being the refractive index of the sample. When a biological object is placed under the microscope objective in the object arm, the light reflected by the reference mirror interferes with the light reflected or backscattered by the sample structures contained in a limited volume that is a slice orthogonal to the objective axis, located at a depth inside the object defined by an optical path length difference of zero. The signal is extracted from the background of incoherent backscattered light using phase-shifting method.

Taking "en face" images allows an easy comparison with histological sections. The resolution, pixel number and sampling requirements results in a field of view that is limited to about 1mm$^2$. Histology sections are typically ~1 cm$^2$ and a pathologist typically evaluates tissue at several magnifications where field of view varies from cm down to µm, in order to make a reliable diagnosis. For this reason, we move the sample on a high precision mechanical platform and we stitch a number of images to display a significant field of view (49).The FF-OCT technology is utilized to build a compact setup (fig 1) that is about the size of a standard optical microscope (310x310x800mm L x W x H) and for this study, was located in the laboratory room used for frozen section analysis (Figure 1B). It is quiet and does not create any unwanted heating, lighting or sound in the room.

*Imaging protocol:*

Written consent was obtained prior to imaging according to the standard procedure in the hospital Tenon (Paris) for each patient undergoing biopsy or surgical resection.

*Sample selection:*

For each patient we imaged at least 4 tissue samples, each averaging 1-1.5cm$^2$: one sample from the tumor, one at the edge of the tumor, and 2 samples from adjacent normal appearing parts of the specimen. Gross examination alone was used to identify the tumor. The size of the tissue samples in this study is considered comparable to the size of those typically acquired and used for frozen section analysis.

*Sample preparation*

The sample tissue was placed in a sample holder and immersed in a saline solution. A specially designed cover slip was carefully placed on the specimen in order to create a flat surface and to reduce optical aberrations. The specimen was raised and gently pushed against the cover slip by displacing the sample holder piston. The holder was then placed under the immersion objective of the microscope. Optical oil was added on top of the cover slip to ensure a continuous optical path.

The specimen is then protected from the environment and is isolated from the air and the objective immersion liquid.

*Image acquisition:*

After the FF-OCT imaging procedure, the samples were fixed in formalin and subjected to routine processing including paraffin embedding, sectioning and HES staining at the same estimated depths than the imaging depth. For consistency, our comparison between images acquired by FF-OCT and histology preparations has been primarily performed with images taken at 20µm below the cover slip, since this depth corresponds to the least attenuated signal on the FF-OCT images and the most representative sections of the paraffin embedded block. Nevertheless, as noted above, signal from breast specimens is detected up to 150µm for the fibrous part and up to 200µm for the fat one.

All FF-OCT images were taken using a non-inversed linear look up table (LUT) scale. Thus, the hypo-scattering structures appear black, whereas the hyper-scattering structures appear white.

In most cases, in order to facilitate the comparison with HES stained slides, the entire tissue sample is imaged. The resulting slides were then scanned to produce whole slide images for use as comparative images.

***Study protocol:***

75 breast specimens were imaged from 22 patients (21 women, 1 man) with a mean age of 58 (range 25-83). A total of 193 images were acquired. In most cases, samples were selected from freshly excised and fixed radical mastectomy specimens, e.g. large malignant or multifocal tumors, benign macroscopic lesions and/or normal appearing tissue present in other areas of the same breast.

Images on fresh tissue offer a better contrast, and all figures in this manuscript are from fresh specimens. Nevertheless images from fixed specimens were easily interpretable by the pathologists and we included images from both fresh and fixed specimen in the sensitivity and specificity study. The samples included normal appearing tissues as well as a large range of breast lesions from men, post- and pre-menopausal women. No discrimination has been made regarding the patient's age or health status, therefore allowing for inclusion the broadest selection of breast tissue and verification that malignancy assessment could be made on any type of tissue.

In most cases, samples were selected from freshly excised radical mastectomy specimens, e.g. large malignant or multifocal tumors, benign macroscopic lesions and/or normal appearing tissue present in other areas of the same breast. A selection of the cases to be sampled was made in order to be able to image a larger amount of different lesions with different macroscopic appearance (nodular or stellate) and different histological types of invasive carcinoma. This information was available from pre-operative diagnoses (mammography and needle core biopsy). Most patients had more than one diagnosis classification associated with their histopathological assessment.

The study was conducted in three phases:

Phase 1: The safety of FF-OCT on human breast tissue was evaluated. Samples were imaged and then processed for routine standard histopathology. The specimens imaged by FF-OCT were assessed for any adverse impact on tissue architecture revealed by HES staining or on immunophenotype assessed by immunochemistry.

Phase 2: The goal of this phase was to highlight the main architectural features of human breast tissue in our images, by comparing these images with standard histopathology slides prepared from the same specimens. This phase facilitated the delineation of the main architectural features seen in human breast tissue, which can be utilized to interpret and characterize the normal/benign and malignant nature of the lesions.

Phase 3: Pathologists were blinded to final histological diagnosis, and were asked to determine based on the FF-OCT image features alone if the specimen was normal/benign or malignant. They receive no macroscopic images of the samples and did not review the histological slides or whole slide images produced from them. The test was conducted long after surgery to prevent any possible familiarity with the case, and the pathologists had no pre-operative assessment information for any of the patients.

**Results:**

Final histopathological diagnoses included 3 cases of invasive lobular carcinoma, 7 cases of invasive ductal adenocarcinoma, 6 cases of lobular carcinoma *in situ*, 6 cases of ductal carcinoma *in situ*, 1 case of invasive mucinous adenocarcinoma and 1 case of ductal hyperplasia.

*Phase 1: Safety*

Samples were imaged using FF-OCT under conditions used in phases 2 and 3 described below.

Six additional radical mastectomy specimens were included in this part of the study. After imaging procedure, tissue architecture was assessed by gold standard histopathology and compared to histological sections corresponding to other specimens of the same region of the breast that did not follow the imaging procedure.

In the case of carcinoma specimens, usual markers tested by immunohistochemistry (IHC) involved in tumor evaluation were quantified in order to verify that FF-OCT imaging procedure does not alter the parameters usually tested. Thus, the signal of Estrogen and Progesterone receptors, Cerb B2 (Her2/neu), ki-67 and E-Cadherin was evaluated and quantified on tumor samples that were imaged with FF-OCT and compared to tumor samples from the same patient that did not follow the imaging procedure.

In the case of normal breast tissue specimen, the signal of other IHC markers such as cytokeratin(CK) 7, 14, 18, 5/6 and vimentin was evaluated and quantified on samples imaged with FF-OCT. The signal obtained was compared to their usual expected signal on normal breast tissue.

After comparison, the value of the different markers for evaluation of the tumourous specimens imaged with FF-OCT were the same as those obtained with matched control specimens from similar regions of the specimen that were not imaged with FF-OCT prior to processing.

For normal breast tissue, the localization of the signal of the different types of CK and vimentin was the same as the expected signal on normal breast tissue.

Therefore, we have shown that no alterations in any of the parameters assessed (histopathology and immunohistochemistry) were observed due to FF-OCT procedure.

The total acquisition and image rendering time for a 1,5cm² breast specimen is 7 minutes. Sample positioning in the sample holder and initialization of the device is 5 minutes long. Together, these make up to the complete imaging procedure requiring only 12 minutes, significantly shorter than what is typically required for frozen section analysis (20-30 min).

*Phase 2 : Correlation of the morphologic features in FF-OCT images with gold standard histopathology*

*Part 1 : Breast tissue characteristics recognition*

After safety assessment of this technique, we focused on distinguishing morphologic features expected in breast tissue on FF-OCT images. This analysis also provided study pathologists with an opportunity to familiarize themselves with the images. We focused on morphological features that were distinguishable in whole slide images of conventional histological preparations and FF-OCT images.

By conducting a comparative analysis with normal breast tissue, we (MA and BSZ) were able to recognize and characterize morphologic features of the major components of breast tissue in the FF-OCT images, including fibroadipose tissue, epithelial ducts, vasculature and fibrous tissue (Figure 2). For example, we found that the galactophorous ducts, when cut tangentially, are recognizable by light grey color, which we hypothesize is generated by their highly scattering thick elastic membranes (Figure 2C). In contrast, the same glands, when cut longitudinally, exhibit a characteristic epithelial layer of varying thickness that appears dark grey (Figure 2B). Intraluminal secretions are sometimes identified. The lobules (Figure 2A) are identified as dark grey granular rounded structures. The vessels (Figure 2D), when cut tangentially, do not have the thick epithelial layer of the galactophorous ducts, but are also associated with the presence of an elastic membrane. The adipocytes (Figure 2E) are hypo-scattering, and appear as black rounded structures. Their membranes produce more scattering and appear grey. The characteristic honeycomb configuration of the fat cells is easily identified. The healthy fibrous tissue appears grainy with medium back-scattering signal (Figure 2G). The fibrous tissue in the stroma is made of hyper-scattering trabeculae (Figure 2H). The scar fibrous tissue is made of thick and large trabeculae, less scattering than the stroma. Calcifications appear white (Figure 2C) due to their very high backscattering levels.

The high resolution of the FF-OCT technique enables the visualization of the very thin membranes of the vessels and those of the adipocytes (Figure 2D, 2E).

Our large scale images display most of these structures that define our grey scale: dark equals absence of tissue or adipocytes, white equals fibrous and dense tissue.

*Part 2: Distinction between benign/normal and malignant tissue*

After gaining an understanding of the normal morphologic features of human breast tissue appearance in our images, we next focused on discerning what aberrations of these morphologic features images would facilitate characterization of the tissue as benign/normal or malignant. The analysis produced a distinct set of criteria that needs to be satisfied for accurate distinction between benign/normal and malignant tissue, along with a recommended workflow that should be followed to make the determination. The workflow and criteria are presented as a flowchart in (Figure 3), and briefly outlined below.

1) First, assess the FF-OCT images using the criteria used for low magnification assessment of standard histology slides. These would include features such as the presence or absence of stellate lesions, and the entanglement of the adipose and fibrous tissues.

2) If there is no evidence of tumor at low magnification, it is necessary to verify the presence of intact normal breast tissue structures such as lobules, galactophorous ducts and vessels.

3) If there is no obvious tumor and there is no evidence of intact normal structures, the following criteria for reading the FF-OCT images must be used.

    a) Assess the shrinkage of the adipose tissue, which is characterized by the presence of adipocytes of different sizes at the periphery of an invasive carcinoma. In most cases, at the interface of the lesion, fat cells appear smaller and less rounded.

    b) Search for the presence of grey zones, typically surrounded by white fibrous structures (corresponding to foci of carcinoma surrounded by fibrous tissue).

    c) Assess the color and the morphology of this fibrous tissue. The appearance of the fibrous tissue is different depending on the type of collagen present. In FF-OCT images, the fibrous tissue of malignant tumor-associated stroma is very white, while it is greyer for scar-associated fibrous tissue or normal/benign breast fibrous tissue. Furthermore, the fibrous tissue architecture differs, usually appearing as thin trabeculae in malignant tumor associated stroma, while appearing as thick trabeculae associated with reactive/reparative change (i.e. wound tissue) (Figure 2F). Trabeculae are not seen in normal or benign fibrous tissue (Figures 2F, 2G, 2H).

Figure 4 consists of an image from normal breast tissue of a post-menopausal woman and the comparable whole slide image of the conventional gold-standard histology preparation. The image shows the characteristic structures of breast tissue such as the lobules, galactophorous ducts, adipocytes and a normal fibrous tissue. There is some entanglement of the fibrous tissue and adipose tissue noted.

The fibrous tissue appears grainy with a medium back scattering level. Ducts are cut longitudinally; thus they look dark grey. Digital zooming of the image reveals acini in the lobules.

Next, in Figure 5, two types of invasive carcinoma can be seen: a stellate tumor (Figure 5A-B) and a nodular one (Figure 5F-G). In both lesions, the trabeculae of the highly scattering tumor-associated fibrous tissue can be observed (Figure 5D, 5H). This is contrast to the normal fibrous tissue that appears grainy and produces less scattering (Figure 5I). Furthermore, tumor-associated adipocytes are smaller than the ones outside of the tumor (Figure 5E). In the stellate tumor (Figure 5B), the fibrous tissue is seen invading the adipose tissue. In the nodular tumor (Figure 5G), foci of carcinoma cells appear as grey zones surrounded by the highly scattering tumor-associated fibrous tissue trabeculae (5H). A circular dilated duct with secretion in the lumen is visible in the center of the nodule. The different aspect of fibrous tissue defines the tumor margins on Figure 5G.

Figure 6 shows an invasive ductal adenocarcinoma with an associated ductal carcinoma in-situ (DCIS)

component. Enlarged lobules and ducts filled with DCIS (Figure 6C, 6D) are clearly visible on the image. In addition, the invasive component is characterized by the presence of highly scattering fibrous tissue (Figure 6B) and foci of darker grey carcinoma cells (Figure 6D).

A fibroadenoma is shown in Figure 7D and the enlarged ductules (Figure 7E) characteristics of this benign lesion are easy to distinguish in the FF-OCT image. DCIS is noted in Figure 7A, characterized by enlarged abnormal lobules (Figure 7C) and ducts (Figure 7B), which can be distinguished easily by digital zooming of the image. The acini in the enlarged lobule are clearly visible and the narrow lumen in the duct is indicative of malignancy.

*Phase 3: Assessment of the diagnostic accuracy of FF-OCT*

Using the above-mentioned workflow and criteria set, 70 and 73, respectfully, breast tissue sample images were reviewed by 2 senior-staff breast pathologists (MA and BS). In isolation of the histopathology or medical history, each classified the images into two diagnostic categories: normal/benign or malignant. Following completion, their designations were compared to the diagnosis based on the gold-standard histopathology preparations of the same tissue used to produce the FF-OCT images. The analysis of the first breast pathologist (MA) yielded 28 true positives, 30 true negatives, 3 false negatives and 9 false positives, giving a sensitivity of 90% and a specificity of 77%. The analysis of the second breast pathologist (BSZ) yielded to 30 true positives, 28 true negatives, 1 false negative and 14 false positives giving a sensitivity of 97% and a specificity of 74%.

Note that for each pathologist, a few images (5-MA, 2-BSZ) were considered uninterpretables, primarily due to a non optimal positioning of the sample under the glass plate used with the sample holder, thus resulting in acquired images at a depth at which the whole sample area was not optically sliced and therefore not entirely visible on the resulting image.

**Discussion:**

In this benchmark pilot study, we describe the first pre-clinical evaluation of FF-OCT for ex-vivo human breast tissue analysis and its ability to facilitate interpretation for tissue diagnosis. First, we have shown that the technique is safe, producing no demonstrable alterations that affect interpretation of conventional histology preparations following imaging, or any alteration in the ability to perform immunohistochemistry for a variety of epitopes. Next, using the images produced, we have demonstrated an ability to assess the microstructure of normal human breast tissue including glands, lobules, galactophorous ducts, and adipose tissue. Lastly, we have developed a set of initial criteria and a recommended workflow for interpreting FF-OCT images which can accurately differentiate whether the alterations observed are benign/normal or malignant. In many cases, a specific histopathological diagnosis can be made by the pathologist, such as invasive carcinoma, carcinoma in-situ, and benign fibroadenomas.

FF-OCT is the first ultra-high resolution optical imaging technique that combines the possibility of obtaining breast tissue images over an adequately large area, while producing images of sufficiently high resolution, thereby allowing pathologists to mimic routine histopathology diagnostic workflow by zooming in and out on the images. We have established a diagnostic decision tree, which would allow pathologists to classify tissue as malignant or benign/normal. While the values obtained for sensitivity and specificity could indeed be improved, it is important to note that this diagnostic accuracy assessment was totally blinded to any clinical information or context, and that the breast tissue remains a very complex organ, with numerous potential histopathological entities.

While the current study has only a limited sample size, and concentrates on a single organ system, the lessons acquired from this initial pilot study will help with the design of future evaluations of tissue samples from all the major organ systems.

We propose that in its current stage of development, FF-OCT imaging technique has great potential for several use cases. First, it represents a paradigm shift in the handling of immediate assessment of resected tissue. For the first time, it allows for primary digital acquisition of enough information to allow pathologist to meaningfully interpret the morphologic features of tissue, on fresh tissue in a non-destructive manner. The lack of need for sample preparation and rapid image capture has the potential to significantly reduce the time it takes to provide this interpretation surgeons. For certain types of frozen sections, such as margin assessment, it has potential to diminish the need for a re-excision and improve post-operative prognosis. Because it is non-destructive, it has great potential to far exceed our current ability to perform both immediate and routine assessment of high risk, small biopsy specimens, such as those obtained from central nervous system tumors. In fact, it could be used to carry out immediate or permanent assessments of tissue biopsies in a variety of contexts. For a typical core needle biopsy with a surface area of 0.2 cm², imaging can be accomplished in about 1 minute. This would then allow for additional testing of the tissue, particularly in cases where fresh non-fixed, non-paraffin embedded tissue is required.

In vitro application of this technology may also be evaluated for both clinical and research tissue banking (where tissue is cryopreserved for future use, such as DNA or RNA extraction). In oncology, the diagnosis and treatment of tumors frequently requires molecular diagnostic information (i.e. gene mutations, translocations, amplifications) to inform targeted therapy. The optimization of these techniques on small biopsy samples is crucial for the pathologist, the clinician and the patient, particularly those who have advanced or metastatic, and thus inoperable, cancer. The fact that the FF-OCT technology does not consume the tissue and does not induce alterations is a real advantage.

Moreover, the standard procedures in a typical pathology laboratory (e.g., embedding tissue, sectioning of a paraffin-embedded tissue with the microtome, frozen section preparation with a cryostat) are still not automated. Thus, these procedures are time consuming and require expert technical personnel. However, it is now possible to foresee the pathology laboratory of the near future, where "triaging" of specimens by FF-OCT could significantly streamline the workflow, such that only truly important specimens are paraffin-embedded and processed for gold standard histopathology.

Finally, in vivo application of FF-OCT technology for endoscopic evaluation will be very helpful for clinicians and surgeons. By utilizing FF-OCT endoscopy, the clinician and pathologist can better evaluate the nature of a suspicious lesion, providing better targeting for areas to sample, decreasing non-diagnostic specimens reaching the lab for conventional histopathological assessment. Such applications are foreseeable for cancers of the respiratory tract, the gastrointestinal tract, the urinary tract and the male and female reproductive tracts. During many of these procedures, for example, the surgeon is unable to palpate the organ directly, limiting their ability to appreciate tactile nature of the possible neoplasm. Consequently, we have recently developed a 2mm-diameter rigid probe and are currently obtaining preliminary results on *in vivo* breast tissue imaging with this experimental setup (50).

**Acknowledgments:**

**Grant Support:** This work was supported in part by Institut National du Cancer (INCA/ONCODIAG n°2009-1-PL BIO 16-ESPCI-1) and by the foundation Pierre Gilles de Gennes pour la recherche.

We would like to thank Eolia Flandre for performing histological sections used for this work.

**References:**

1. American Cancer Society : Cancer Facts and Figures; 2011.
2. Silverstein MJ, Recht A, Lagios MD, Allred DC, Harms SE, Holland R et al. Image-detected breast cancer: state-of-the-art diagnosis and treatment. J Am Coll Surg 2009;209:504–20.
3. Cendán JC, Coco D, Copeland EM. Accuracy of intraoperative frozen-section analysis of breast cancer lumpectomy-bed margins. J Am Coll of Surg 2005;201:194-8.
4. Nakazawa H, Rosen P, Lane N, Lattes R. Frozen section experience in 3000 cases: Accuracy, limitations, and value in residence training. Am J Clin Pathol 1968;49:41-51.
5. Olson TP, Harter J, Munoz A, Mahvi DM, Breslin T. Frozen section analysis for intraoperative margin assessment during breast-conserving surgery results in low rates of re-excision and local recurrence. Ann Surg Oncol 2007;14:2953-60.
6. McLaughlin SA ,Ochoa-Frongia LM, Patil SM, Cody HS 3rd, Sclafani LM. Influence of frozen-section analysis of sentinel lymph node and lumpectomy margin status on reoperation rates in patients undergoing breast-conservation therapy. J Am Coll  Surg 2008; 206:76-82.
7. Ishak K. Benign tumors and pseudotumors of the liver. Appl Pathol 1988;6:82-104.
8. Taxy J. Frozen section and the surgical pathologist :  A point of view. Arc Pathol Lab Med 2009;133:1135-8.
9. Sienko A, Allen TC, Zander DS, Cagle PT. Frozen section of lung specimens. Arch Pathol Lab Med 2005;129:1602-9.
10. Nagasue N, Akamizu H, Yukaya H, Yuuki I. Hepatocellular pseudotumor in the cirrhotic liver. Report of three cases. Cancer 1984;54:2487-94.
11. Weinberg E, Cox C, Dupont E,  White L, Ebert M, Greenberg H et al. Local Recurrence in lumpectomy patients after imprints cytology margin evaluation. Am J Surg 2004;188:349-54.
12. Valdes EK, Bollbol SK, Cohen JM, Feldman SM. Intra-operative touch preparation cytology ; does it have a rolen in re-excision lumpectomy? Ann Surg Oncol 2007;14:1045-50.
13. Goldfeder S, Davis D, Cullinan J. Breast specimen radiography : can it predict margin status of excised breast carcinoma? Acad Radiol 2006; 187:339-44.
14. Swanson GP, Rynearson K, Symmonds R. Significance of margins of excision on breast cancer recurrence. Am J Clin Oncol-Cancer 2002;25:438-41.
15. Gonzalez-Angulo AM, Morales-Vasquez  F, Hortobagyi GN. Overview of resistance to systemic therapy in patients with breast cancer. Adv  Exp Med Biol 2007; 608:1-22.
16. Papa MZ, Zippel D, Koller M, Klein E, Chetrit A, Ari GB et al. Positive margins of breast biopsy: is reexcision always necessary? J Surg Oncol 1999;70:167-71.
17. Willner J, Kiricuta IC,  Kölbl O. Locoregional recurrence of breast cancer following mastectomy:

always a fatal event? Results of univariate and multivariate analysis. Int J Radiat Oncol 1997;37:853-63.
18. Cabioglu N, Hunt KK, Sahin AA, Kuerer HM, Babiera GV, Singletary SE et al. Role for intraoperative margin assessment in patients undergoing breast-conserving surgery. An Surg Oncol 2007;14:1458-71.
19. Fleming FJ, Hill ADK, Mc Dermott EW, O'Doherty A, O'Higgins NJ, Quinn CM. Intraoperative margin assessment and re-excision rate in breast conserving surgery. Eur J Surg Oncol 2004;30:233-7.
20. Garreau DS, Patel YG, Li Y, Aranda I, Halper AC, Nehal KS. Confocal mosaicing microscopy in skin excisions: a demonstration of rapid surgical pathology. J Microsc 2009;233:149–159.
21. Masters BR, So PT. Confocal microscopy and multi-photon excitation microscopy of human skin in vivo. Opt Express 2001;8:2-10.
22. Huang D, Swanson EA, Lin CP, Schuman JS, Stinson WG, Chang W, Hee M et al. Optical coherence tomography. Science 1991;254: 1178–81.
23. Dubois A, Boccara AC. Full-field Optical Coherence Microscopy. In: Drexler W, Fujimoto GJ, editors. Optical Coherence Tomography. New York-Basel: Springer; 2007. p.565-91.
24. Dubois A, Boccara AC. Full-field OCT. Med Sci (Paris) 2006;22:859-64.
25. Pawley JB. editors. Handbook of Biological Confocal Microscopy. 3rd ed. Berlin: Springer;2006.
26. Zhou C, Cohen DW, Wang Y, Lee HC, Mondelblatt AE, Tsai TH, Aguirre AD et al. Integrated Optical Coherence Tomography and Microscopy for Ex Vivo Multiscale Evaluation of Human Breast Tissues. Cancer Res 2010;70:10071-79.
27. Swanson EA, Izatt JA, Hee MR, Huang D, Lin CP, Schuman JS et al. In vivo retinal imaging by optical coherence tomography. Opt Lett 1993;18:1864-66.
28. Grieve K, Paques M, Dubois A, Sahel J, Boccara C, Le Gargasson JF. Ocular tissue imaging using ultrahigh-resolution, full-field optical coherence tomography. Invest Ophthalmol Vis Sci 2004;45:4126-31.
29. Drexler W, Morgner U, Ghanta RK, Kärtner FX, Schuman JS, Fujimoto. JG et al. Ultrahigh-resolution ophthalmic optical coherence tomography. Nat Med 2001;7: 502-7.
30. Betz CS, Stepp H, Havel M, Jerjes W, Upile T, Hopper C et al. A set of optical techniques for improving the diagnosis of early upper aerodigestive tract cancer. Med Las Appl 2008;23:175–85.
31. Ozawa N, Sumi Y, Shimozato K, Chong C, Kurabayashi T. In vivo imaging of human labial glands using advanced optical coherence tomography. Oral Surg Oral Med O 2009;108:425-29.
32. Chen Y, Aguirre AD, Hsiung PL, Desai S, Herz PR, Pedrosa M et al. Ultrahigh resolution optical coherence tomography of Barrett's esophagus: preliminary descriptive clinical study correlating images with histology. Endoscopy 2007;39:599-605.
33. Jerjes W, Upile T, Conn B, Betz CS, Abbas S, Jay A et al. Oral leukoplakia and erythroplakia subjected to optical coherence tomography: preliminary results. Brit J Oral Max Surg 2008;46:e7.
34. Tearney GJ, Brezinski ME, Southern JF, Bouma BE, Boppart SA, Fujimoto JG. Optical biopsy in human pancreatobiliary tissue using optical coherence tomography. Dig Dis Sci 1998;43:1193-9.
35. Boppart SA, Luo W, Marks DL, Singletary KW. Optical coherence tomography: feasibility for basic research and image-guided surgery of breast cancer. Breast Cancer Res Treat 2004;84:85-97.
36. Adie SG, Boppart SA. Optical Coherence Tomography for Cancer Detection. In Rosenthal E, Zinn KR Editors. OpticalImaging of Cancer Clinical Applications. New York: Springer 2009:209-50.
37. Hsiung PL, Phatak DR, Chen Y, Aguirre AD, Fujimoto JG, Connolly JL. Benign and malignant lesions in the human breast depicted with ultrahigh resolution and three-dimensional optical coherence

tomography. Radiology 2007; 244:865–74.
38. Nguyen FT, Zysk AM, Chaney EJ, Kotynek JG, Oliphant UJ, Bellafiore FJ et al. Intraoperative Evaluation of Breast Tumor Margins with Optical Coherence Tomography. Cancer Res 2009;69:8790-96.
39. Zysk AM, Boppart SA. Computational methods for analysis of human breast tumor tissue in optical coherence tomography images. J Biomed Opt 2006;11:1-7.
40. Luo W, Nguyen FT, Zysk AM, Ralston TS, Brockenbrough J, Marks DL et al. Optical biopsy of lymph node morphology using optical coherence tomography. Tech Canc Res Treat 2005;4:539-48.
41. Boppart S, Luo W, Marks D, Singletary K. Optical Coherence Tomography: Feasibility for Basic Research and Image-guided Surgery of Breast Cancer. Breast Cancer Res Tr 2004; 84:85-97.
42. Feldichtein FI, Gelikonov VM, Gelikonov GV. Design of OCT scanners. In : Bouma BE, Tearney GJ editors. Handbook of Optical Coherence Tomography. New York- Basel: Marcel Dekker Inc; 2002. p.125-42.
43. Heikal A, Webb W. One –and two-photon time resolved fluorescence spectroscopy of selected fluorescent markers : photobleaching, triplet-, and singlet-state dynamics. Biophys J 1999;76:260a.
44. Jain M, Shukla N, Manzoor M, Nadolny S, Mukherjee S. Modified full field optical coherence tomography : A novel tool for histology of tissues. J Pathol Inform 2011; 2:28.
45. Dubois A, Moneron G, Grieve K, Boccara AC. Three-dimensional cellular level imaging using full-field optical coherence tomography. Phys Med Biol 2004;49:1227-34.
46. Dubois A, Grieve K, Moneron G, Lecaque R, Vabre L, Boccara AC. Ultra high resolution full-field optical coherence tomography. Appl Opt 2004;43:2874-83.
47. Binding J, Ben Arous J, Léger JF, Gigan S, Boccara AC, Bourdieu L. Brain refractive index measured in vivo with high-NA defocus-corrected full-field OCT and consequences for two-photon microscopy. Opt Express 2011; 19:4833-47.
48. Dubois A, Vabre L, Boccara AC, Beaurepaire E. High-resolution full-field optical coherence tomography with a Linnik microscope. Appl Optics 2002;41: 805–12.
49. Beck J, Murray J, Dennis Willows AO, Cooper MS. Computer assisted visualizations of neural networks: expanding the field of view using seamless confocal montaging. J Neurosci Meth 2000; 98:155-63.
50. Latrive A, Boccara AC. In vivo and in situ cellular imaging full-field optical coherence tomography with a rigid endoscopic probe. Biomed Opt Express.2011;2:2897-904.

Figures:

**Figure 1 :**

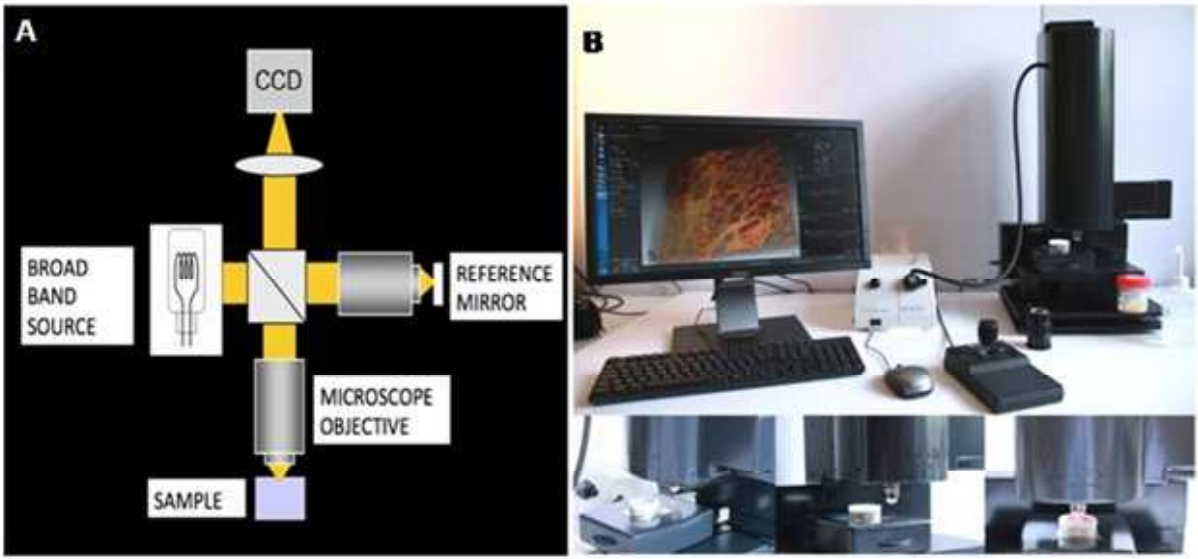

**Figure 1 : FF-OCT experimental setup based on a Linnik interferometer configuration (1A). Compact set-up used for the study (1B)**

**Figure 2 :**

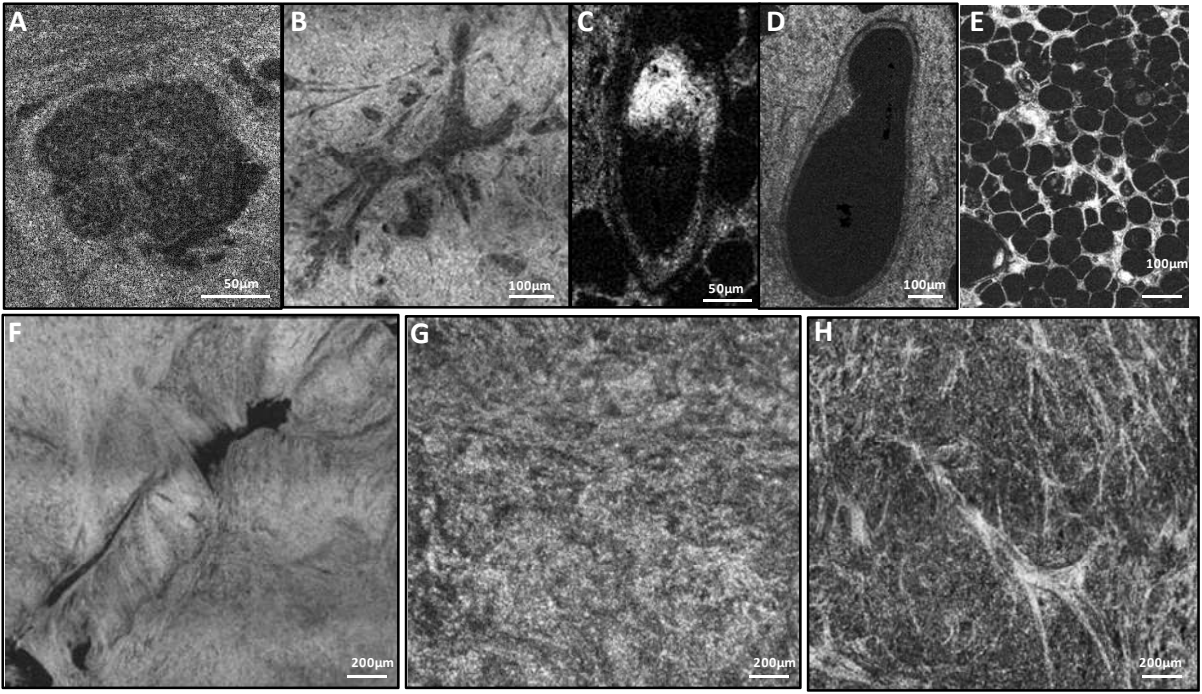

**Figure 2 : Breast tissue basic structures: lobule(A), galactophorous duct (B), cross section of a galactophorous duct with calcifications (C), vessel (D), adipocytes (E). Scar fibrous tissue (F), normal fibrous tissue (G), fibrous tissue surrounding carcinomatous cells in tumourous stroma (H). Images were acquired at 20µm beneath the tissue surface.**

**Figure 3** :

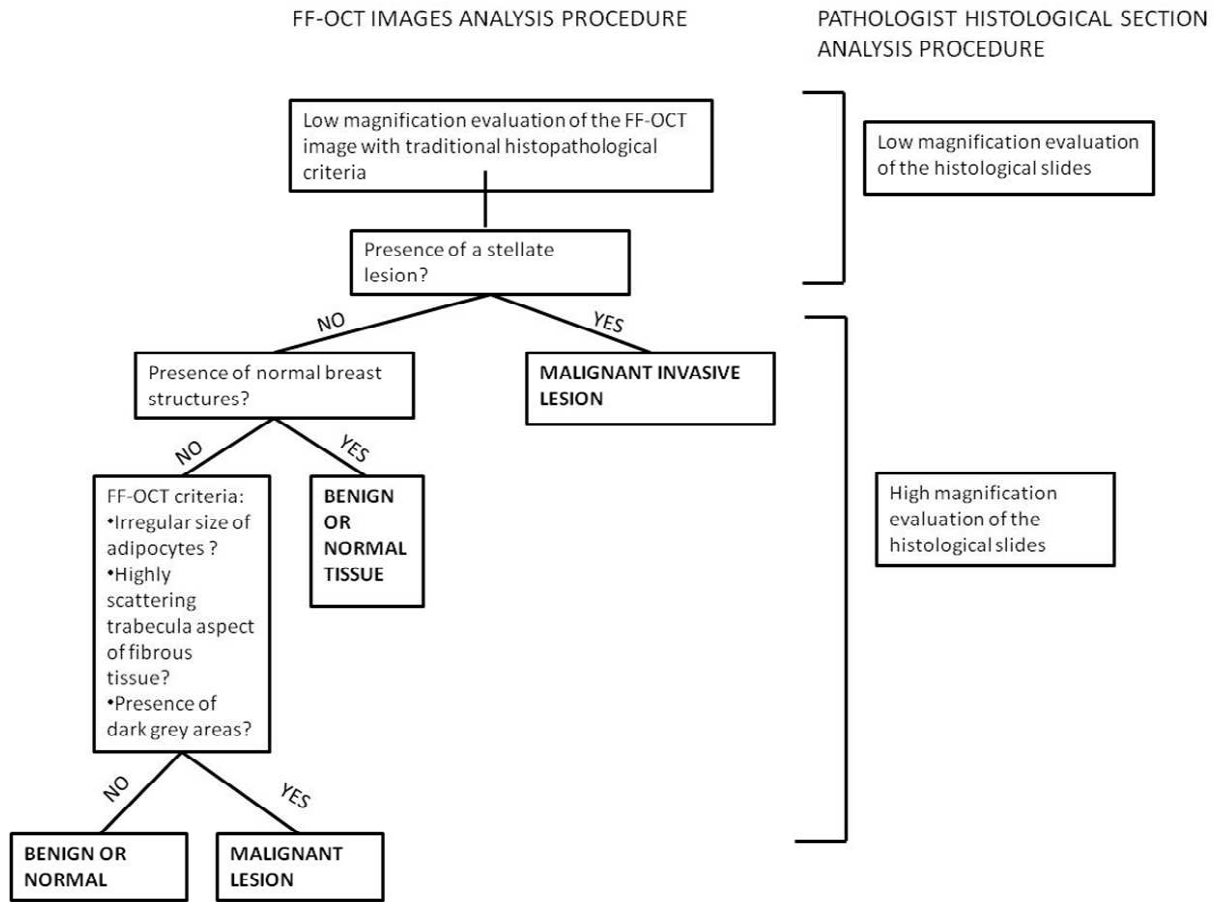

Figure 3 : Diagnosis decision tree for FF-OCT images on human breast tissue

**Figure 4** :

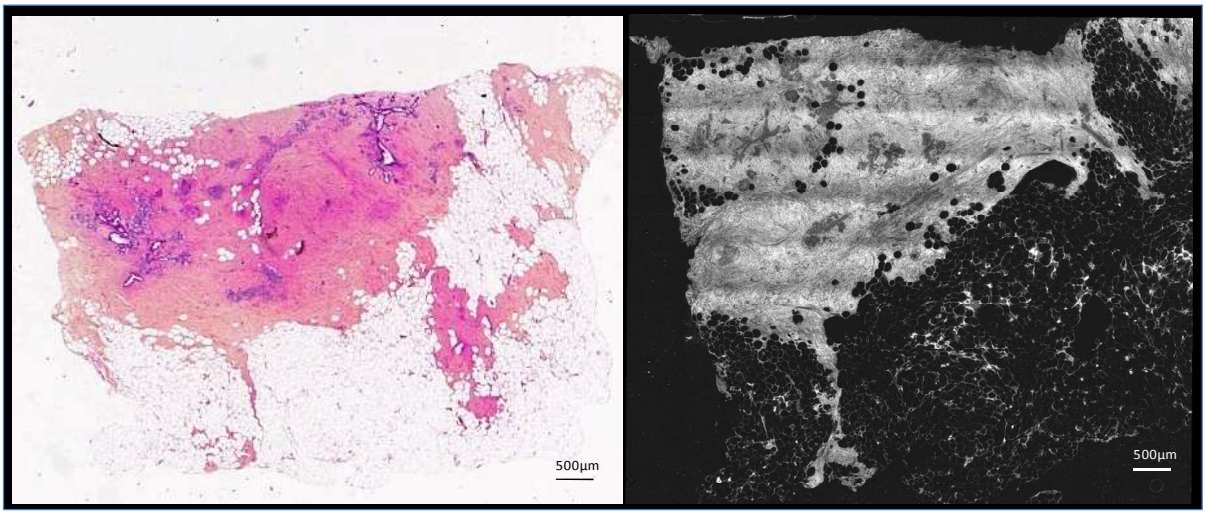

Figure 4 : Healthy breast tissue specimen. FF-OCT image was acquired at 20μm beneath the tissue surface.

**Figure 5 :**

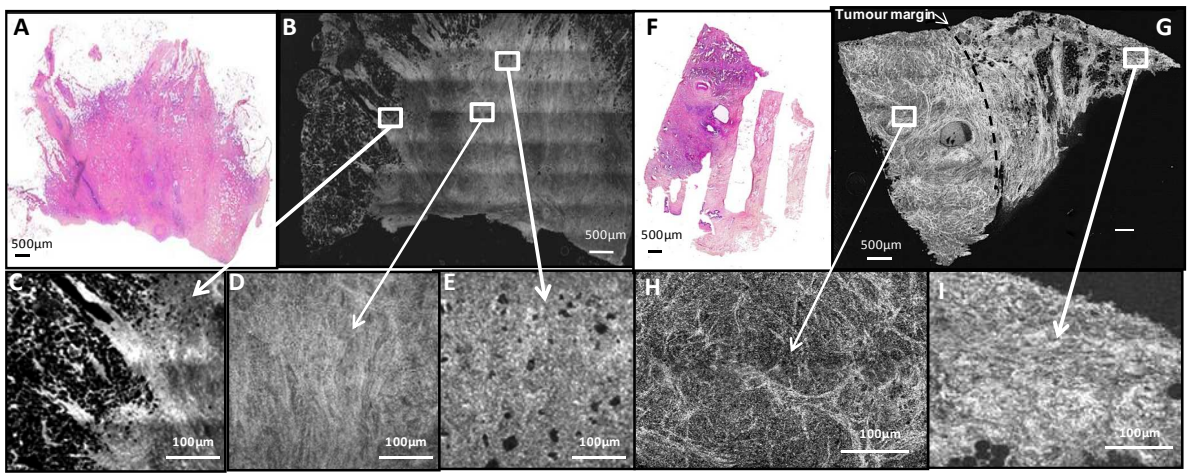

**Figure 5 : Different types of invasive adenocarcinoma : stellate (A) and nodular (B). FF-OCT Images were acquired at 20µm beneath the tissue surface**

**Figure 6 :**

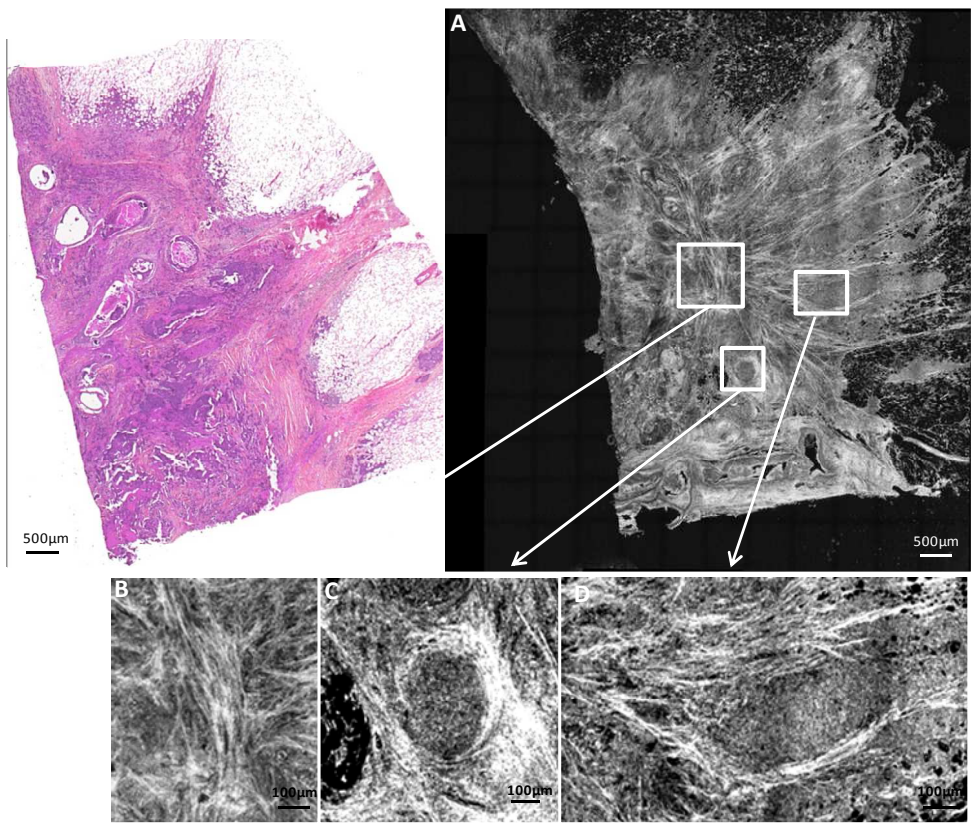

**Figure 6 : Ductal invasive adenocarcinoma with ductal Carcinoma in situ (DCIS) component. FF-OCT Images were acquired at 20µm beneath the tissue surface**

**Figure 7** :

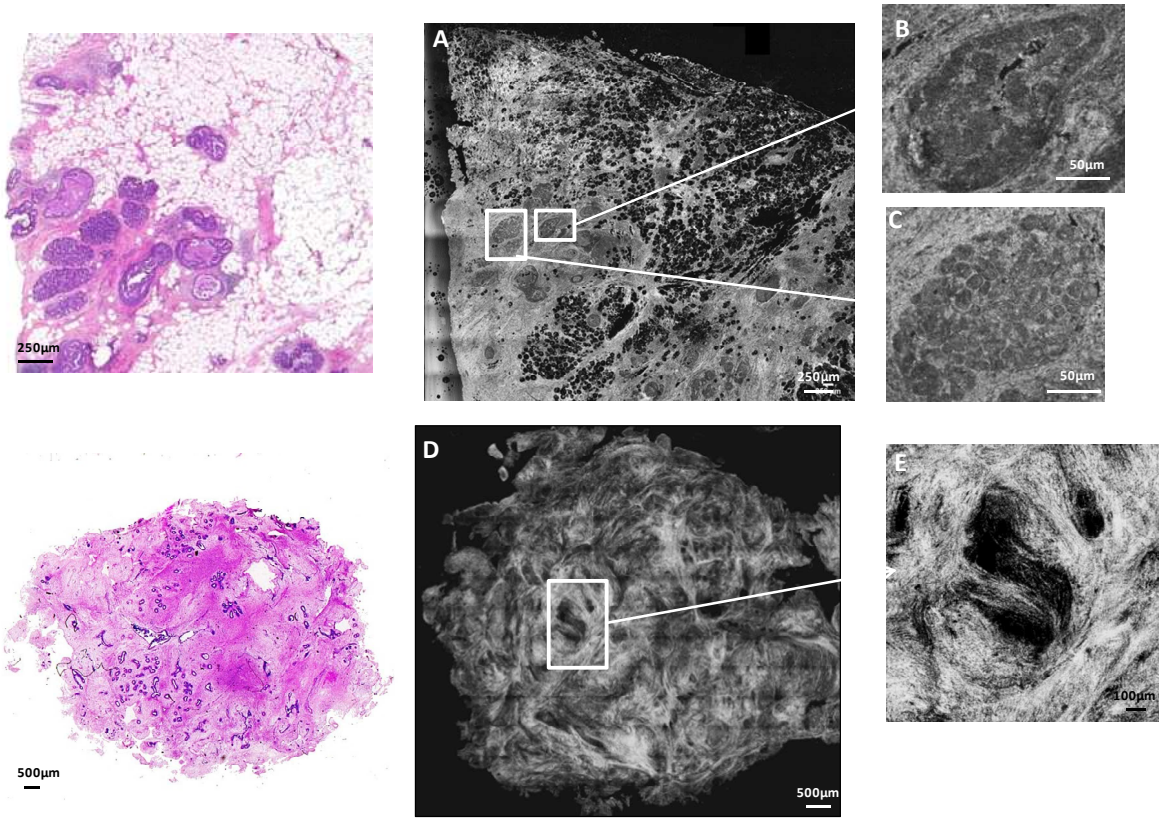

**Figure 7 : Ductal carcinoma in situ (A) with an abnormal duct (B) and an enlarged lobule (C). Fibroadenoma (D) and an enlarged ductule (E). FF-OCT images were acquired at 20µm beneath the tissue surface.**